# Thermodynamic model of twisted bilayer graphene: Configuration entropy matters

Weidong Yan[a], Langquan Shui[a], Wengen Ouyang[a], Ze Liu[a,b,c*]

[a] *Department of Engineering Mechanics, School of Civil Engineering, Wuhan University, Wuhan, Hubei 430072, China*

[b] *State Key Laboratory of Water Resources & Hydropower Engineering Science, Wuhan University, Wuhan 430072, China*

[c] *The Institute of Technological Science, Wuhan University, Wuhan, Hubei 430072, China*

*Corresponding authors. E-mail addresses: ze.liu@whu.edu.cn (Z. L.)

**Abstract:** Twisted bilayer materials have attracted tremendous attention due to their unique and novel properties. Here, we derive a thermodynamic model for twisted bilayer graphene (tBLG) within the framework of the classical statistical mechanics, based on which, the configuration entropy reflecting the number of micro-status in moiré unit-cells, is directly derived from the Helmholtz free energy with a clear physical interpretation. More importantly, we show the configuration entropy of tBLG relative to the AB-stacked bilayer graphene is proportional to the logarithmic function of the ratio of moiré period ($a_m$) and the atomic lattice constant ($a$) as $S_{\mathrm{tBLG}} - S_{\mathrm{AB}} = 24k_B \ln\left(\frac{a_m}{a}\right)$, which we found dominates the Helmholtz free energy of tBLG and can well explain experimental observations in superlubric contacts. Our work provides a theoretical foundation for studying moiré effect of incommensurate contact interfaces and could facilitate twisting based applications such as superlubricity.

**Keywords:** Twisted bilayer graphene; Moiré pattern; Configuration entropy; Superlubricity;

## 1 Introduction

Tuning the electronic, mechanical and frictional properties of layered materials by changing the interlayer twist angle has attracted extensive attention in recent years (Gao et al., 2018; Huang et al., 2021; Koren et al., 2016; Liu et al., 2012a; Wang et al., 2019; Weston et al., 2020; Zhang et al., 2020a; Zhang et al., 2012; Zhang et al., 2020b). For instance, by increasing the twist angle of contacted layered materials (such as graphene/graphite, graphene/*h*-BN, $MoS_2/MoS_2$), the friction drops drastically to superlubric states (Dienwiebel et al., 2004; Dietzel et al., 2013; Feng et al., 2013; Filippov et al., 2008; Hirano et al., 1997; Hod et al., 2018; He et al., 2022; Koren et al., 2016; Martin et al., 1993; Song et al., 2018; Wang et al., 2016a; Yaniv and Koren, 2019). The underlying mechanism mainly attributes to the lateral force cancellation due to the incommensurate registry (Dienwiebel et al., 2004; Hod, 2013; Verhoeven et al., 2004). Besides, the out-of-plane deformation regulated by the moiré pattern within the layered materials (Guo et al., 2011; Mandelli et al., 2019; Ouyang et al., 2016; Song et al., 2018) also contributes significantly to friction force. The reason is that the energy is mainly dissipated through the out-of-plane deformation channel due to ultralow bending rigidity of layered materials.

However, there is still lacking a clear description of the above phenomenon on the basis of fundamental physical principles. One possible way to fill this gap is to study the thermomechanical properties of layered materials based on statistical mechanics, as is successfully applied to the suspend and supported monolayer graphene (Fasolino et al., 2007; Gao and Huang, 2014; Los et al., 2009; Wang et al., 2016b). The main limitation of this method is that it treats the graphene layer as a continue film and neglects its lattice effect, thus it cannot consider the moiré pattern formed at the interface, which is found to dominate the twist angle based physical properties of vertically stacked layered materials (Cao et al., 2018; Huang et al., 2021; Koren et al., 2016; Wang et al., 2016a; Wang et al., 2019; Weston et al., 2020; Zhang et al., 2020b).

In this work, we developed a theoretical model to include the effect of moiré pattern within the framework of the classical statistical mechanics, based on which the origin of interlayer friction in tBLG is discussed. Interestingly, we found that even neglecting the effect of out-of-plane deformation, the interlayer friction cannot be zero because of the twist angle dependent configuration entropy, which we directly derived from the Helmholtz free energy and clarified its physical interpretation based on the Boltzmann entropy. The friction of superlubric contact originating from the configuration entropy is further formulated as $\tau_S = 72k_BT/(\alpha L^3)\cdot\cot(\theta/2)$, where $k_B$ is the Boltzmann constant, $T$ is the absolute temperature, $L$ is the size of the graphene flake, $\theta$ is the twist angle and $\alpha = \sqrt{2} + \mathrm{artanh}\frac{\pi}{8}$ for a square graphene flake.

## 2 Thermodynamic model of tBLG

*2.1 Analysis on the moiré-dependent deformation*

The model system of tBLG is shown in Fig. 1a, we choose square membrane here, the periodic boundary condition is applied in both $x$ and $y$ directions (Ouyang et al., 2020), and the fixed substrate is applied to neglect the effect of substrate stiffness (Zhang et al., 2015). By rotating the top graphene with respect to the fixed bottom graphene with a twist angle $\theta$, moiré pattern (or moiré superlattice) with a period of $a_m$ forms (Fig. 1b). Mathematically, the geometrical moiré patterns are always periodic (regardless of the arrangement of atoms) and its periodicity $a_m(\theta) = \frac{a}{2\sin\frac{\theta}{2}}$ (Bistritzer and MacDonald, 2011; Rong and Kuiper, 1993), where $a$ is the lattice constant of graphene.

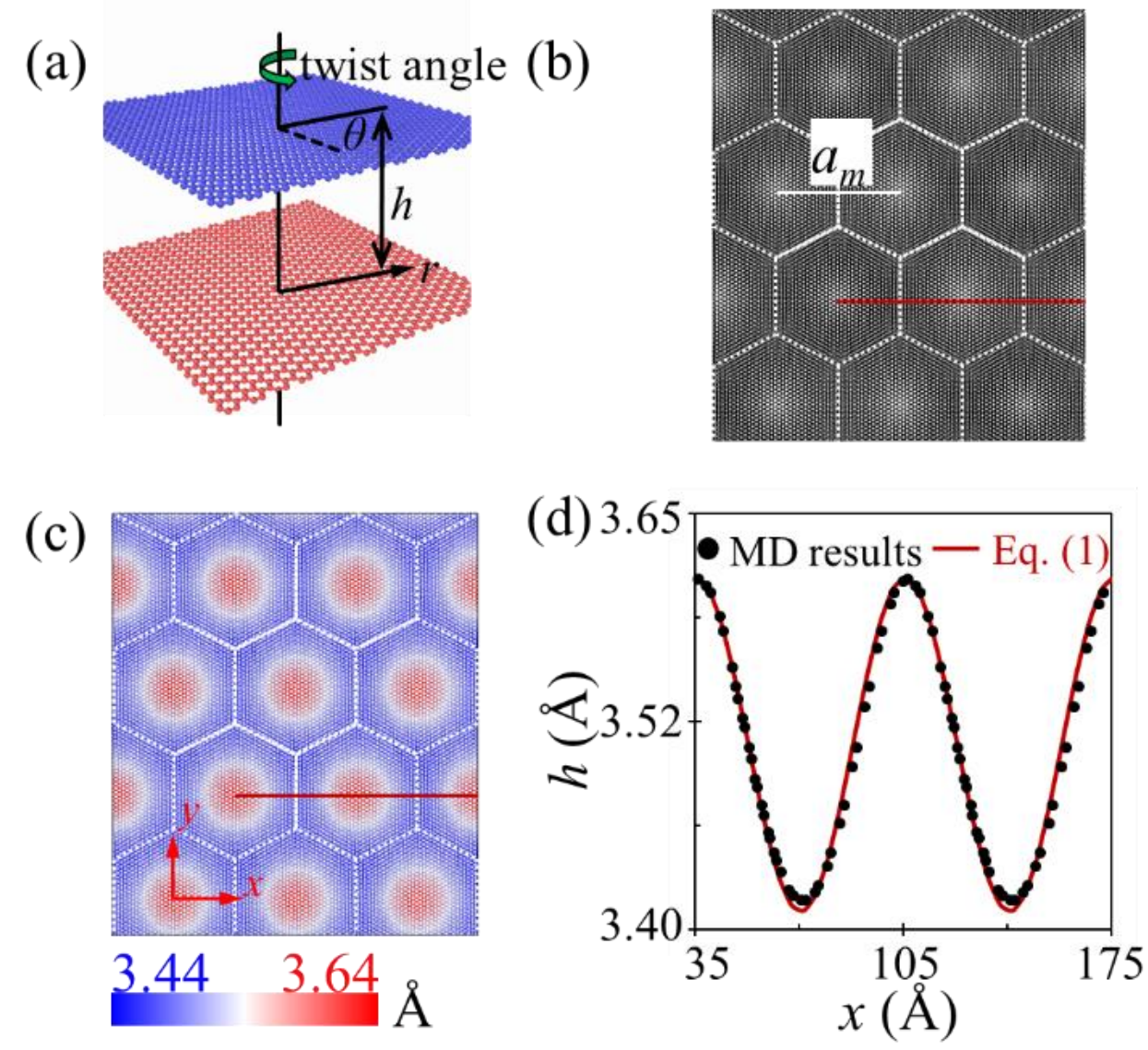

**Fig. 1.** (a) Model illustration of twisted bilayer graphene (tBLG), where the periodic boundary condition is applied in lateral directions. (b) Typical moiré pattern for tBLG with a twist angle of 2°, where the white hexagons represent the cells of the geometric moiré pattern and $a_m$ is its periodicity. (c) The atomic height distribution of the top graphene in (b) after energy optimization at 0 K, where the bottom graphene was fixed. (d) The height profile of the top graphene (a) along the red line in panels (b) and (c), the black line in panel (d) is obtained by fitting the height profile along $y$ = 12.18 nm with Eq. (1).

Due to the extremely low bending rigidity of 2D materials, out-of-plane deformation and thermal rippling can be easily observed in monolayer graphene and influence its mechanical properties (Xu and Buehler, 2010; Ahmadpoor et al., 2017; Chang et al., 2019; Gao and Huang, 2014; Guo et al., 2016; Yang et al., 2021). To show the effect of moiré superlattice on its out-of-plane deformation, we performed geometry optimization for tBLG presented in Fig. 1a, REBO potential and Kolmogorov–Crespi (KC) potential (Kolmogorov and Crespi, 2005) with a refined parameterization (Ouyang et al., 2018) were used for the intralayer and interlayer interaction, respectively. The geometry optimization was performed with FIRE algorithm with a force criteria of $10^{-6}$ eV/Å. All simulations are performed using LAMMPS (Plimpton, 1995). The optimized configuration is presented in Fig. 1c, which shows the out-of-plane deformation of the top graphene exactly follows the moiré supercells. The moiré-dependent deformation mode induced by the interlayer potential energy at 0 K can be described by the cosine function, similar result has been reported before (Steele, 1973; Wijk et al., 2015; Zhao, 2020), and such a deformation pattern has been shown to be stable even at room temperature (Ouyang et al., 2021). Hence, we used the following formula to describe the moiré-dependent deformation:

$$h^{0\mathrm{K}} \approx h_{\mathrm{eq}}^{0\mathrm{K}} + \Delta h^{0\mathrm{K}} \cos py \left(\cos py + \cos\sqrt{3}px\right) = h_{\mathrm{eq}}^{0\mathrm{K}} + \frac{\Delta h^{0\mathrm{K}}}{2} + \sum_{k=1}^{6} \frac{\Delta h^{0\mathrm{K}}}{4} \exp(i\boldsymbol{p}_k \cdot \boldsymbol{r}) \quad (1)$$

where $h_{\mathrm{eq}}^{0\mathrm{K}}$ and $\Delta h^{0\mathrm{K}}$ are the equilibrium distance and the amplitude of the out-of-plane deformation at 0 K, respectively. $\boldsymbol{r} = (x, y)$ is the position vector, $\boldsymbol{p}_k$ $(k = 1, 2, 3, 4, 5, 6)$ are vectors that defines a moiré supercell:

$$\boldsymbol{p}_k = p\left(\sin\frac{k\pi}{3}, \cos\frac{k\pi}{3}\right) \quad (2)$$

where $p = 2\pi/(\sqrt{3}a_m)$, $a_m$ is the size of moiré pattern. For AB-stacked bilayer graphene, the periodicity should be replaced with the lattice constant of graphene: $\boldsymbol{p}_k^{\boldsymbol{0}} = p^0\left(\sin\frac{k\pi}{3}, \cos\frac{k\pi}{3}\right)$, where $p^0 = 2\pi/(\sqrt{3}a)$. We found that Eq. (2) can well describe both atomic-dependent deformation mode for AB-stacked bilayer graphene and the moiré-dependent deformation mode for tBLG with $\theta > 0°$. Fitting the height profile extracted from the MD simulation (along the red line in Fig. 1c) with Eq. (1) shows excellent agreement (red line in Fig. 1d). The fitting of the entire surface can be found in Fig. 2.

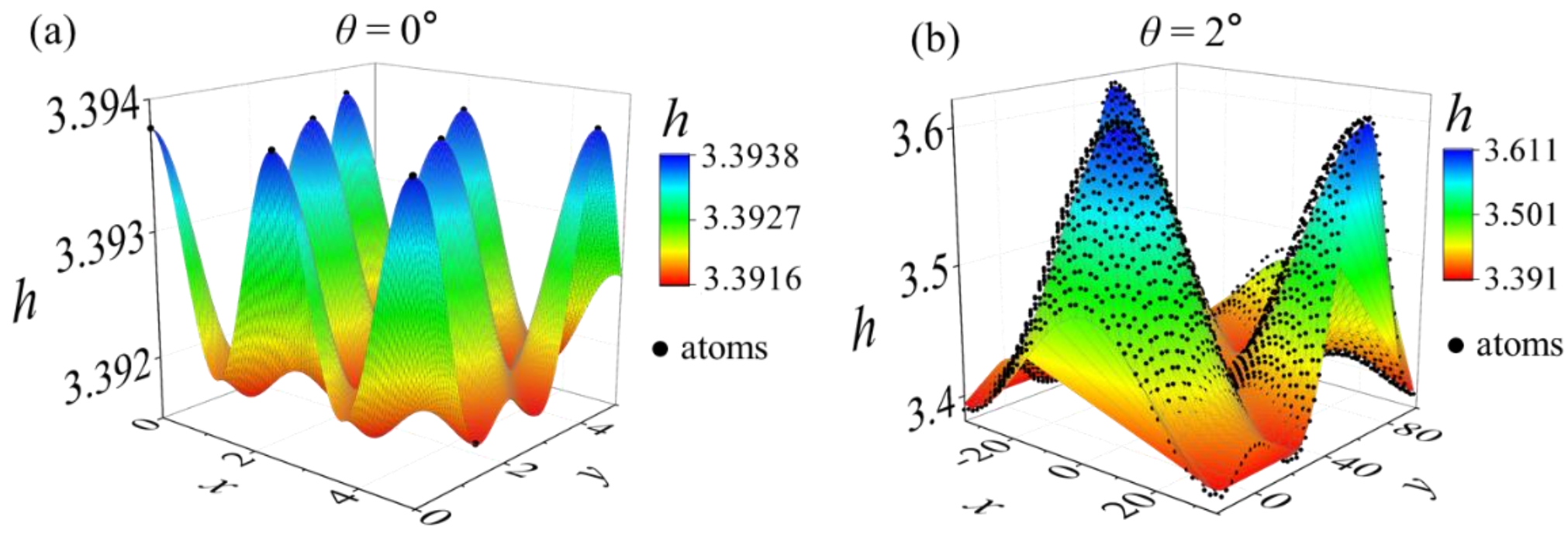

**Fig. 2.** Fitting the moiré-dependent deformation mode of tBLG with a twist angle of 0° (a) and 2° (b) using Eq. (1). The black dots represent the positions of atoms in the top graphene as shown in Fig. 1. The colored surface is obtained by fitting with Eq. (1). The parameters in Eq. (1) were extracted from MD simulations: $h_{\mathrm{eq}}^{0\mathrm{K}} = 3.39$ Å, $\Delta h^{0\mathrm{K}} = 9.11 \times 10^{-4}$ Å, $a_m = 2.46$ Å for $\theta = 0°$, and $h_{\mathrm{eq}}^{0\mathrm{K}} = 3.41$ Å, $\Delta h^{0\mathrm{K}} = 0.10$ Å, $a_m = 70.32$ Å for $\theta = 2°$.

For finite temperature, the thermal motion of atoms induces further out-of-plane deformation in the top graphene, here we assume that the total out-of-plane deformation can be superimposed the thermal induced out-of-plane deformation ($W(\mathbf{r})$) to the moiré-dependent deformation mode. Mathematically, $W(\mathbf{r})$ can be generally expressed in the form of Fourier series, we adopted the following widely used formula (Ahmadpoor et al., 2017; Chang et al., 2019; Gao and Huang, 2014; Nelson and Peliti, 1987; Plimpton, 1995; Wang et al., 2016b):

$$W(\boldsymbol{r}) = \sum_{(n,l)\in\boldsymbol{\Omega}} w_{nl} \exp(i\boldsymbol{q}_{nl}\cdot\boldsymbol{r}) + \sum_{k=1}^{6} b_k \exp(i\boldsymbol{p}_k\cdot\boldsymbol{r}) \tag{3}$$

where the two terms at the right of Eq. (3) correspond to the lattice deformation modes and the moiré-dependent deformation mode, respectively. $\boldsymbol{q}_{nl} = \frac{2\pi}{L}(n,l)$ is the wave vector in the 2D space, $\boldsymbol{\Omega} = \left\{(n,l)\middle| n\in\mathbb{Z}, l\in\mathbb{Z}, n^2+l^2 \le \frac{L^2}{a^2}\right\}$, $w_{nl} = \frac{1}{L^2}\iint_{\mathbb{S}} W(x,y)e^{-i\boldsymbol{q}_{nl}\cdot\boldsymbol{r}}\mathrm{d}x\mathrm{d}y$ are the corresponding Fourier coefficients, $b_k$ are the corresponding coefficients of thermal fluctuations originating from the moiré-dependent deformation mode. For sufficiently low temperature, the out-of-plane deformation of the top graphene can be expressed as:

$$h(x,y) = h^{0\mathrm{K}} + W = h_{\mathrm{eq}}^{0\mathrm{K}}\left(\widehat{h^{0\mathrm{K}}} + \widehat{W}\right) \tag{4}$$

where $\widehat{h^{0\mathrm{K}}} = h^{0\mathrm{K}}/h_{\mathrm{eq}}^{0\mathrm{K}}$ and $\widehat{W} = W/h_{\mathrm{eq}}^{0\mathrm{K}}$.

*2.2 Helmholtz free energy of tBLG*

Based on the classical statistical mechanics, the probability density function (PDF, $\rho$) of each deformation mode is given by the Boltzmann distribution at thermal equilibrium, $\rho = \exp(-E/k_B T)/Z$, where $E$ is the total potential energy of each configuration, $k_B$ is the Boltzmann constant and $Z$ is the partition function (Ahmadpoor et al., 2017; Chang et al., 2019; Wang et al., 2016b).

The total potential energy contains two parts, the bending energy and interlayer potential energy. Because the magnitude of interlayer shear modulus is several orders smaller than that of the in-plane elastic modulus, the in-plane strain is ignored in the following analysis. Hence, the potential energy of the system consists of two parts, one is the interlayer potential and the other is the bending energy of the top graphene. Firstly, the bending deformation energy of the top graphene (with a size of $L\times L$) can be calculated as:

$$U = \frac{D\left(h_{eq}^{0\mathrm{K}}\right)^2}{2}\int_{-\frac{L}{2}}^{\frac{L}{2}}\int_{-\frac{L}{2}}^{\frac{L}{2}}\left[\frac{\partial^2\left(\widehat{h^{0\mathrm{K}}}+\widehat{W}\right)}{\partial x^2} + \frac{\partial^2\left(\widehat{h^{0\mathrm{K}}}+\widehat{W}\right)}{\partial y^2}\right]^2 \mathrm{d}x\mathrm{d}y \tag{5}$$

where $D$ is the bending rigidity of graphene. Substituting Eqs. (1)- (4) into Eq. (5), we obtain:

$$
\begin{aligned}
U &= \frac{3D}{16}(\Delta h^{0\mathrm{K}})^2 p^4 L^2 \left[1 + \frac{1}{3pL}\left(4\sin\frac{pL}{2} + \frac{4}{\sqrt{3}}\sin\frac{\sqrt{3}pL}{2} + \sin pL\right)\right. \\
&\left. + \frac{16}{3\sqrt{3}p^2L^2}\left(\frac{8}{3}\cos\frac{pL}{4} + \cos\frac{\sqrt{3}pL}{4}\right)\sin\frac{pL}{2}\sin\frac{\sqrt{3}pL}{4}\right] \\
&+ \sum_{k=1}\frac{D}{2}L^2p^4\left(\frac{\Delta h^{0\mathrm{K}}}{4}\bar{w}_k + \frac{\Delta h^{0\mathrm{K}}}{4}w_k + \frac{\Delta h^{0\mathrm{K}}}{4}\bar{b}_k + \frac{\Delta h^{0\mathrm{K}}}{4}b_k + w_k\bar{b}_k + b_k\bar{w}_k\right) + \sum_{(n,l)\in\boldsymbol{\Omega}}\frac{D}{2}L^2\,|w_{nl}|^2|\boldsymbol{q}_{nl}|^4 \\
&+ \sum_{k=1}\frac{D}{2}L^2|b_k|^2p^4 \qquad (6)
\end{aligned}
$$

For large enough graphene flake (for instance $L \sim 100\ \mathrm{nm}$ ), $pL = \frac{2\pi L}{\sqrt{3}a_m} \sim 12.8 \gg 1$, here $a_m = 28.18\ \mathrm{nm}$ is the size of moiré period for tBLG with a twist angle $\theta = 0.5°$. Noting that the typical sample size is several microns in experiments (Liu et al., 2012a), $pL \gg 1$ is always satisfied and Eq. (6) can thus be approximated as:

$$
\begin{aligned}
U \approx \frac{3D}{16}(\Delta h^{0\mathrm{K}})^2 p^4 L^2 &+ \sum_{k=1}\frac{D}{2}L^2p^4\left(\frac{\Delta h^{0\mathrm{K}}}{4}\bar{w}_k + \frac{\Delta h^{0\mathrm{K}}}{4}w_k + \frac{\Delta h^{0\mathrm{K}}}{4}b_k + \frac{\Delta h^{0\mathrm{K}}}{4}\bar{b}_k + w_kb_k + b_k\bar{w}_k\right) \\
&+ \sum_{(n,l)\in\boldsymbol{\Omega}}\frac{D}{2}L^2\,|w_{nl}|^2|\boldsymbol{q}_{nl}|^4 + \sum_{k=1}\frac{D}{2}L^2|b_k|^2p^4 \qquad (7)
\end{aligned}
$$

On the other hand, the total interaction energy between the top graphene and the bottom graphene substrate can be obtained as:

$$
\begin{aligned}
V &= \rho_s \int_{-\frac{L}{2}}^{\frac{L}{2}}\int_{-\frac{L}{2}}^{\frac{L}{2}} u[h(x,y)]\,\mathrm{d}x\mathrm{d}y \\
&\approx \rho_s \int_{-L/2}^{L/2}\int_{-L/2}^{L/2} u\left(h_{\mathrm{eq}}^{0\mathrm{K}}\right) + u'\left(h_{\mathrm{eq}}^{0\mathrm{K}}\right)\left(h - h_{\mathrm{eq}}^{0\mathrm{K}}\right) + \frac{1}{2}u''\left(h_{\mathrm{eq}}^{0\mathrm{K}}\right)\left(h - h_{\mathrm{eq}}^{0\mathrm{K}}\right)^2\,\mathrm{d}x\mathrm{d}y \qquad (8)
\end{aligned}
$$

where $u[h(x,y)]$ is the interaction energy between a carbon atom in the top graphene with the bottom graphene substrate. $h_{\mathrm{eq}}^{0\mathrm{K}}$ is the equilibrium distance at absolute zero temperature, and $u''\left(h_{\mathrm{eq}}^{0\mathrm{K}}\right)$ is the second order derivative coefficient in the Taylor expansion of the interaction energy $u$. Substituting Eqs. (1)- (4) into Eq. (8), we have:

$$
\begin{aligned}
V &= \rho_s L^2 u\left(h_{\mathrm{eq}}^{0\mathrm{K}}\right) \\
&\quad + \frac{1}{2}\rho_s u''\left(h_{\mathrm{eq}}^{0\mathrm{K}}\right)\left\{\frac{1}{4}(\Delta h^{0\mathrm{K}})^2 L^2 + \frac{2(\Delta h^{0\mathrm{K}})^2}{3p^2}\sin\frac{Lp}{4}\left(3Lp\cos\frac{Lp}{4} + 8\sqrt{3}\sin\frac{\sqrt{3}Lp}{4}\right)\right. \\
&\quad + \frac{(\Delta h^{0\mathrm{K}})^2}{72p^2}\left[9Lp\left(3Lp + 4\sin\frac{Lp}{2} + \sin Lp\right) + 64\sqrt{3}\sin\frac{\sqrt{3}Lp}{4}\left(3\sin\frac{Lp}{4} + \sin\frac{3Lp}{4}\right)\right. \\
&\quad \left. + 12\sqrt{3}\sin\frac{\sqrt{3}Lp}{2}\left(Lp + 2\sin\frac{3Lp}{2}\right)\right] + \sum_{k=1}^{6}\frac{\Delta h^{0\mathrm{K}}}{4}L^2\bar{w}_k + \sum_{k=1}^{6}\frac{\Delta h^{0\mathrm{K}}}{4}L^2\bar{b}_k \\
&\quad + \sum_{k=1}^{6}\frac{\Delta h^{0\mathrm{K}}}{4}L^2 w_k + \sum_{k=1}^{6} w_k\bar{b}_k L^2 + \sum_{k=1}^{6}\frac{\Delta h^{0\mathrm{K}}}{4} b_k L^2 + \sum_{k=1}^{6}\bar{w}_k b_k L^2 + \sum_{k=1}^{6} L^2 b_k^2 \\
&\quad \left. + L^2\sum_{(n,l)\in\mathbf{\Omega}}|w_{nl}|^2\right\}
\end{aligned}
\tag{9}
$$

Similarly, in the case of $pL \gg 1$, Eq. (9) simplifies to:

$$
\begin{aligned}
V &\approx \rho_s L^2 u\left(h_{\mathrm{eq}}^{0\mathrm{K}}\right) \\
&\quad + \frac{1}{2}\rho_s u''\left(h_{\mathrm{eq}}^{0\mathrm{K}}\right)L^2\left[\frac{5}{8}(\Delta h^{0\mathrm{K}})^2 + \frac{\Delta h^{0\mathrm{K}}}{4}\sum_{(n,l)\in\mathbf{\Omega_\theta}}(\bar{w}_{nl} + w_{nl}) + \sum_{(n,l)\in\mathbf{\Omega}}|w_{nl}|^2\right. \\
&\quad \left. + \sum_{k=1}^{6}\frac{\Delta h^{0\mathrm{K}}}{4}\left(b_k + \bar{b}_k\right) + \sum_{k=1}^{6}\left(b_k\bar{w}_k + \bar{b}_k w_k\right) + \sum_{k=1}^{6} b_k^2\right]
\end{aligned}
\tag{10}
$$

According to the statistical mechanics, the partition function $Z$ in probability density function reads:

$$
Z = \int_{\mathbb{R}^{|\Omega|}} \exp(-\beta E)\prod_{(n,l)\in\mathbf{\Omega}}\mathrm{d}(\mathrm{Re}\widehat{w}_{nl})\prod_{(n,l)\in\mathbf{\Omega}}\mathrm{d}(\mathrm{Im}\widehat{w}_{nl})\prod_{k=1}^{6}\mathrm{d}\left(\mathrm{Re}\hat{b}_k\right)\prod_{k=1}^{6}\mathrm{d}\left(\mathrm{Im}\hat{b}_k\right)
\tag{11}
$$

where $E = U + V$, $\beta = 1/k_B T$.

Substituting Eqs. (6)- (9) into Eq. (11), we obtain:

$$
\begin{aligned}
&Z \\
&= \exp\left[-\beta L^2\left(\rho_s u\left(h_{\mathrm{eq}}^{0\mathrm{K}}\right) + \frac{1}{8}(\Delta h^{0\mathrm{K}})^2\rho_s u''\left(h_{\mathrm{eq}}^{0\mathrm{K}}\right)\right)\right] \\
&\times \prod_{k=1}^{6}\frac{4\pi}{\sqrt{3}\left(h_{\mathrm{eq}}^{0\mathrm{K}}\right)^2\beta L^2\left[D|\boldsymbol{p}_k|^4 + \rho_s u''\left(h_{\mathrm{eq}}^{0\mathrm{K}}\right)\right]}\prod_{(n,l)\in\mathbf{\Omega}}\frac{2\pi}{\left(h_{\mathrm{eq}}^{0\mathrm{K}}\right)^2\beta L^2\left[D|\boldsymbol{q}_{nl}|^4 + \rho_s u''\left(h_{\mathrm{eq}}^{0\mathrm{K}}\right)\right]}
\end{aligned}
\tag{12}
$$

Based on Eq. (12), the Helmholtz free energy of a tBLG can be calculated as:

$$
\begin{aligned}
F &= -\frac{1}{\beta}\ln Z \triangleq F_1 + F_2 + F_3 \\
&\qquad = \rho_s L^2 u\left(h_{\text{eq}}^{0\text{K}}\right) + \frac{1}{8}(\Delta h^{0\text{K}})^2 L^2 \rho_s u''\left(h_{\text{eq}}^{0\text{K}}\right) \\
&\qquad -\frac{1}{\beta}\left[\sum_{k=1}^{6} \ln \frac{4\pi}{\sqrt{3}\left(h_{\text{eq}}^{0\text{K}}\right)^2 \beta L^2 \left[D|\boldsymbol{p}_k|^4 + \rho_s u''\left(h_{\text{eq}}^{0\text{K}}\right)\right]}\right. \\
&\qquad \left. + \sum_{(n,l)\in\boldsymbol{\Omega}} \ln \frac{2\pi}{\left(h_{\text{eq}}^{0\text{K}}\right)^2 \beta L^2 \left[D|\boldsymbol{q}_{nl}|^4 + \rho_s u''\left(h_{\text{eq}}^{0\text{K}}\right)\right]}\right] \qquad (13)
\end{aligned}
$$

It is noted that the first term in Eq. (13) is the interlayer potential energy, the second term is induced by the initial out-of-plane deformation at 0 K, and the third term represents the effect of thermal fluctuations. We found that the term of $\rho_s u''\left(h_{\text{eq}}^{0\text{K}}\right)$ has a minor effect on the third term (Fig. 3e), then Eq. (13) can be simplified as follows:

$$
\begin{aligned}
F &\cong \rho_s L^2 u\left(h_{\text{eq}}^{0\text{K}}\right) + \frac{1}{8}(\Delta h^{0\text{K}})^2 L^2 \rho_s u''\left(h_{\text{eq}}^{0\text{K}}\right) \\
&\qquad -\frac{1}{\beta}\left[\sum_{k=1}^{6} \ln \frac{4\pi}{\sqrt{3}\left(h_{\text{eq}}^{0\text{K}}\right)^2 D\beta L^2 |\boldsymbol{p}_k|^4} + \sum_{(n,l)\in\boldsymbol{\Omega}} \ln \frac{2\pi}{\left(h_{\text{eq}}^{0\text{K}}\right)^2 D\beta L^2 |\boldsymbol{q}_{nl}|^4}\right] \qquad (14)
\end{aligned}
$$

*2.3 Numerical calculation of the Helmholtz free energy*

In this section, numerical calculations of the free energy based on Eq. (14) are presented. To obtain the required parameters in Eq. (14), the stable configuration of tBLG with different twisted angles were firstly simulated by the MD simulation (Fig. 3). The initial geometry configurations were generated using the same method as detailed in Ref. (Ouyang et al., 2020). The geometry optimization was performed with FIRE algorithm with a force criteria of $10^{-6}$ eV/Å. After geometry optimization, the interlayer potential energy of tBLG and the out-of-plane deformation of the top graphene as a function of twist angle ($\theta$) were obtained (Figs. 3a- b). To calculate $u''\left(h_{\text{eq}}^{0\text{K}}\right)$, one needs to know how the interlayer potential energy changes with respect to the interlayer distance. Taking typical tBLG with a twist angle of $1.12°$ as an example ($h_{\text{eq}}^{0\text{K}} = 3.41$ Å), by shifting the top graphene of the optimized tBLG rigidly from $h_{\text{eq}}^{0\text{K}} - 0.5$ Å to $h_{\text{eq}}^{0\text{K}} + 0.5$ Å with a step of 0.002 Å, we obtained the relationship between the potential energy $u$ and the interlayer distance $h$ (Fig. 3c). Here, we focus on the change of $u$ near $h_{\text{eq}}^{0\text{K}}$, where the interlayer potential energy versus the interlayer distance can be approximated by the quadratic function and $u''\left(h_{\text{eq}}^{0\text{K}}\right)$ can be readily determined by fitting (the insert of Fig. 3c) Similarly, $u''\left(h_{\text{eq}}^{0\text{K}}\right)$ for other twisted angles can also be calculated (Fig. 3d). In addition, when calculating the Helmholtz free energy, we found that the effect of the term $u''\left(h_{\text{eq}}^{0\text{K}}\right)$ is negligible (Fig. 3e).

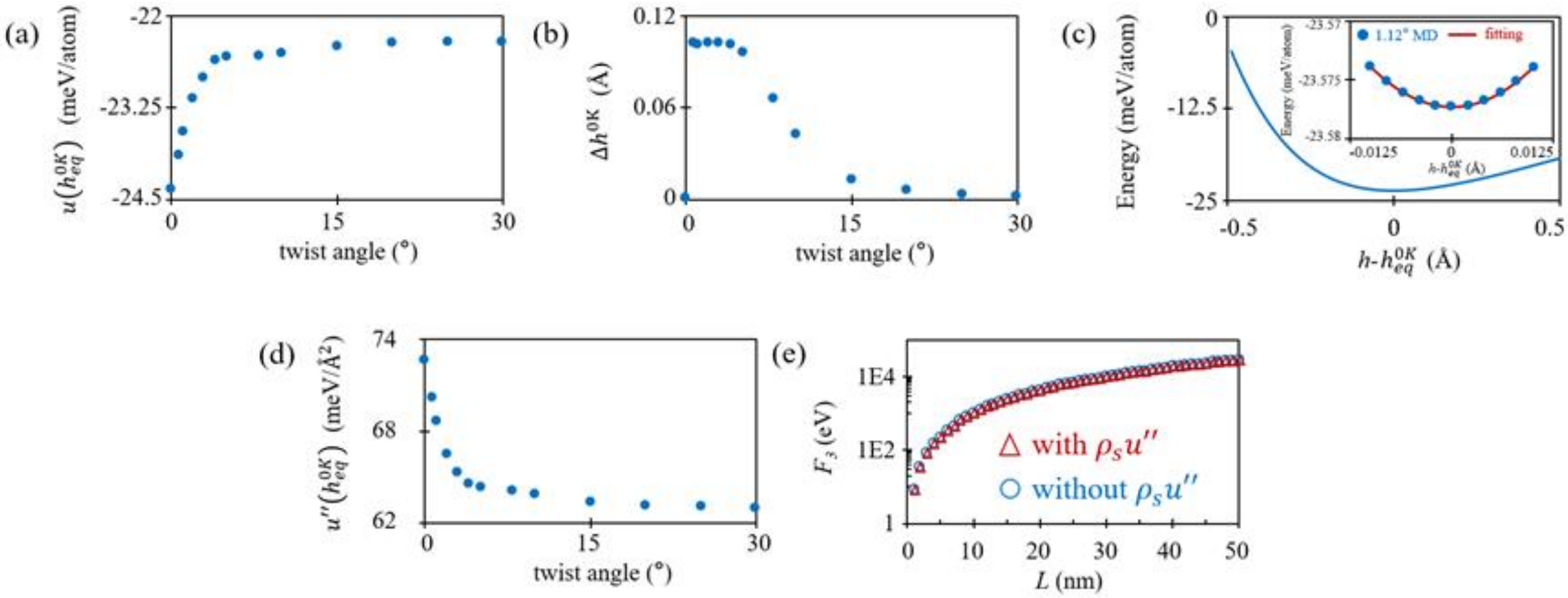

**Fig. 3.** Numerical calculation of the Helmholtz free energy. (a)-(b) The dependence of the interlayer potential energy and the out-of-plane deformation amplitude of the top graphene on the twisted angle. (c) The relation between the interlayer potential and the interlayer distance. The insert shows the fitting of the data with a quadratic function. (d) The fitted $u''\left(h_{\mathrm{eq}}^{0\mathrm{K}}\right)$ as a function of $\theta$. (e) Calculated $F_3$ for the tBLG ($\theta = 1.12°$) with and without the term $\rho_s u''\left(h_{\mathrm{eq}}^{0\mathrm{K}}\right)$, respectively. The parameters used in the calculation are: $h_{\mathrm{eq}}^{0\mathrm{K}} = 3.41$ Å, $u''\left(h_{\mathrm{eq}}^{0\mathrm{K}}\right) = 68.66\ \mathrm{meV/Å^2}$, $\rho_s = 0.38\ 1/\mathrm{Å^2}$, $\Delta h^{0\mathrm{K}} = 0.10$ Å, $T = 300$ K.

If taking the Helmholtz free energy of the tBLG at the AB-stacked mode as the reference, then we have:

$$\Delta F \triangleq \Delta F_1 + \Delta F_2 + \Delta F_3 = \rho_s L^2 \Delta u + \frac{1}{8} L^2 \rho_s u''\left(h_{\mathrm{eq}}^{0\mathrm{K}}\right)\left[(\Delta h^{0\mathrm{K}})^2 - \left(\Delta h_{AB}^{0\mathrm{K}}\right)^2\right] - \frac{24}{\beta} \ln \frac{a_m(\theta)}{a} \tag{15}$$

where, $\Delta F_1$, $\Delta F_2$ and $\Delta F_3$ correspond to the contributions of the interlayer potential energy, the initial out-of-plane deformation at 0 K, and the configuration entropy (as we will clarify later) to the Helmholtz free energy, respectively. $\Delta u = u\left(h_{\mathrm{eq}}^{0\mathrm{K}}\right) - u_{AB}$ is the interlayer potential energy difference between a tBLG and the AB-stacked bilayer graphene, $a_m$ is the size of moiré supercell of tBLG and $a$ is the lattice constant of graphene.

Typical results for a sample size of 7 nm are shown in Fig. 4, we found that during the quasi-static rotation, the second term is negligible (Fig. 4b) compared to other terms, and both $\Delta F_1$ and $\Delta F_3$ decrease as the twisted angle decreases but the former takes the minimum value at $\theta = 0°$ and the latter takes the maximum at $\theta = 0°$ (Figs. 2a and c), that is, $\Delta F_1$ and $\Delta F_3$ dominate for $\theta = 0°$ and $\theta > 0°$, respectively. The finding predicts that if a graphene flake is released from an incommensurate contact with fixed bottom graphene or a thick graphite substrate, it tends to rotate to the AB-stacked state spontaneously, which agrees well with experimental observations (Filippov et al., 2008; Liu et al., 2012b; Wang et al., 2016a). Note that the potential energy curve quickly flattens after deviating from the AB-stacked state (i.e. almost independent on the twist angle for $\theta > 5°$, Fig. 4a), especially for large size of top graphene flake (Zhang and Tadmor, 2017). We attribute the observed spontaneously rotation motion to the contribution of entropy energy since it is dominant compared to other terms (Fig. 4). Physically, the spontaneous rotation from an incommensurate contact to the AB-stacked state corresponds to an entropy increasing process.

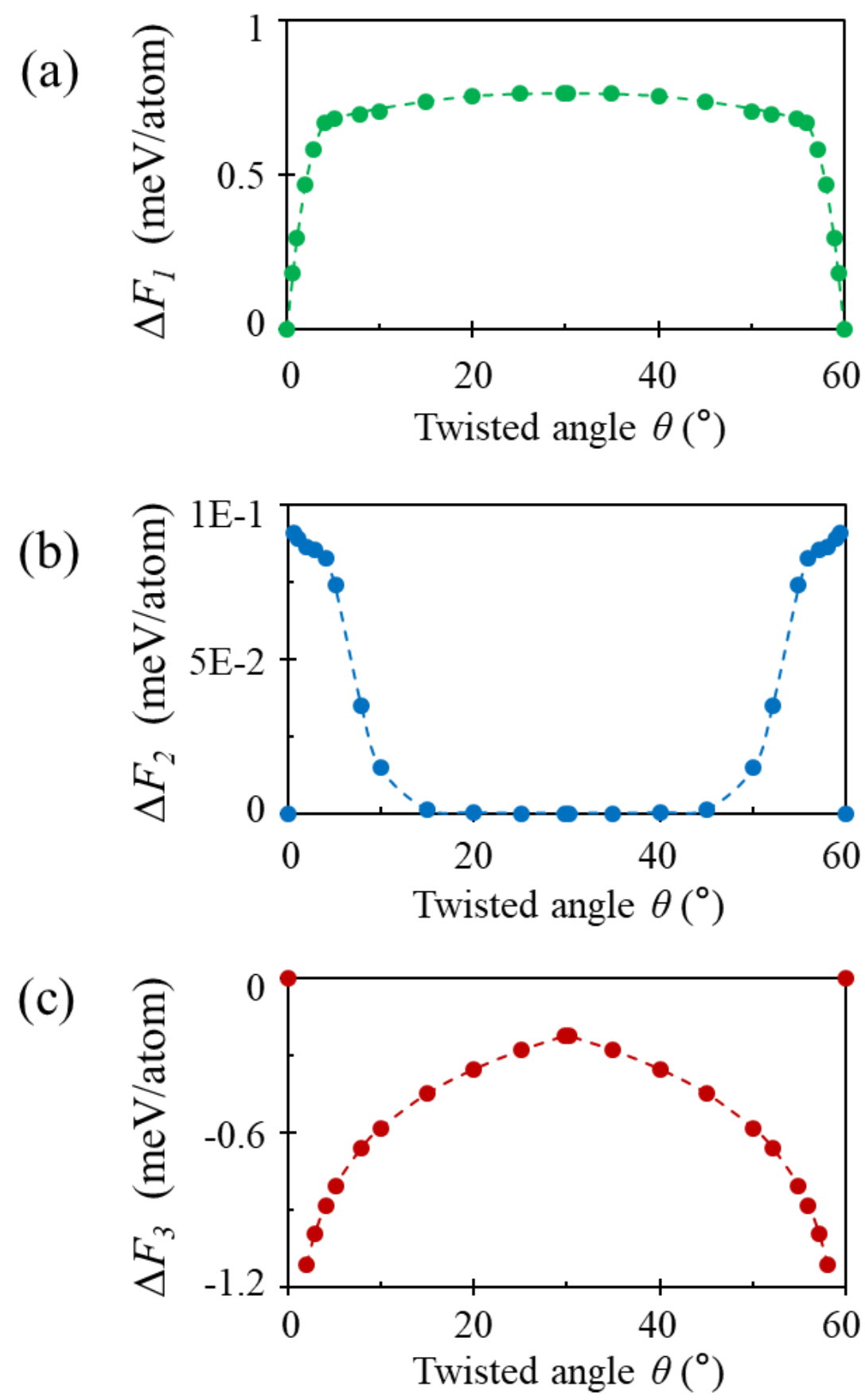

**Fig. 4.** (a)-(c) Contribution to the Helmholtz free energy from $\Delta F_1$ (a), $\Delta F_2$ (b), and $\Delta F_3$ (c). The results were calculated based on Eq. (15), where the parameters are obtained from MD simulations.

*2.4 The configuration entropy of tBLG*

To get the physical interpretation of the third term in Eq. (15), we first noted that $\Delta F_3$ is formally similar to the Boltzmann's entropy formula, meanwhile, once the top graphene flake is rotated out of the AB-stacked state ($\theta = 0°$) with the fixed bottom graphene, the regular arrangement of atoms between the top and bottom graphene will be broken and the disorder appears. Quantifying such a disorder requires the calculation of the number of micro-status of the system under the condition of the same system potential energy. To get insights into the relationship between the moiré pattern and the system potential energy, we output the potential energy of each atom in the top layer graphene that interacting with the bottom layer graphene (Fig. 5a), where the twist angle is 5.086°. We found that the potential energy distribution of all atoms in each moiré unit-cell (black hexagons in Fig. 5a) is similar. This feature becomes clearer if the twist angle is big so that the number of atoms in one moiré unit-cell is small. Typical results are shown in Fig. 5b for $\theta$ = 21.7869° (commensurate supercells) and $\theta$ = 21.7500° (incommensurate supercells), respectively, where we projected the potential energy of all atoms in the top layer graphene (contains 47532 atoms) into the polar coordinate-potential energy ($r$-$u$) plane. It is clear that for the bilayer graphene with a twist angle of 21.7869°, there are 14 energy levels with an almost equal number of atoms (3390, 3391, and 3401 atoms, respectively) in each of the energy levels (left of Fig. 5b), this number of energy levels exactly equals to the number of atoms in one moiré unit-cell. Once rotating the bilayer graphene a little bit to a twist angle of $\theta$ = 21.7500°, the 14 energy levels become 14 energy bands but keeping the same energy level distribution (right of Fig. 5b).

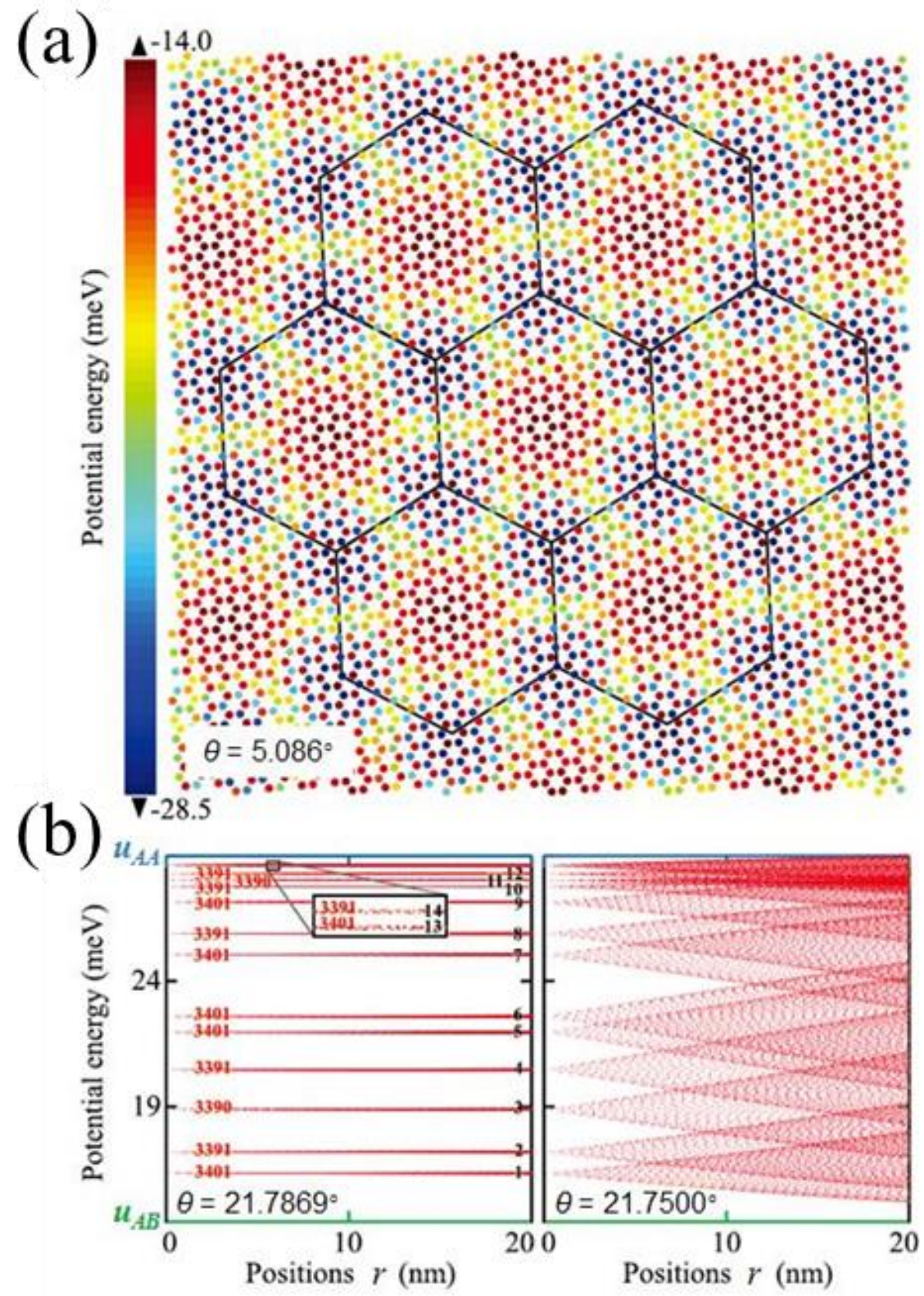

**Fig. 5.** Correlation of the moiré pattern with the system potential energy. (a) Potential energy distribution of all atoms in the top layer graphene of tBLG, where the moiré pattern (black hexagons) is also given for reference. The radius of the upper and bottom graphene is set as 20 nm and 22.5 nm, respectively. The twist angle is 5.086°. (b) Projection of the potential energy of all atoms in the top layer graphene into the energy-polar coordinate ($r$-$u$) plane. The twist angles in left and right of (b) are 21.7869° and 21.7500°, respectively.

The above observations lead us to conclude that the number of micro-status of tBLG is directly proportional to the area of one moiré unit-cell. To illustrate how we correlate the micro-status of a twisted bilayer graphene (with twist angle $\theta$) with the area of one unit-cell of the moiré pattern, we take one commensurate superlattice ($\theta$ = 21.78°) as an example (Fig. 6). Essentially, it requires to find out all of the micro-status (or configurations) under the conditions of:

(i) Keep the twist angle unchanged.

(ii) The total energy of each micro-status must be the same.

Noting that the each of atoms within one moiré unit-cell possesses different interaction energy with the bottom graphene substrate due to their different position vectors (Fig. 5), such differences are reflected in the moiré patterns (magenta diamonds in Fig. 6), which make all the atoms of the upper graphene be distinguishable with the moiré pattern as reference, e.g. the labeled atoms A, B, and A', B' in Fig. 6a. As required by the condition (i), if we translationally move the upper graphene from a location of one labeled atom (e.g. the location of atom A in Fig. 6a) to all the positions of rest atoms within the same moiré unit-cell (e.g. the position of atom B in Fig. 6a), the moiré patterns do not change (Fig. 6b) and so do the system energy (as required by the condition (ii)), however, it is clear that the positions of the labeled atoms in Fig. 6(a) change relative to the moiré patterns (Fig. 6b). Therefore, we conclude that Fig. 6a and b represent two micro-status of the bilayer graphene system with the same potential energy. On the other hand, if we translationally move the upper graphene from a location of one labeled atom (e.g. the location of atom A in Fig. 6a) to all the positions of atoms located at other moiré unit-cells (e.g. the position of atom A' in Fig. 6a), neither the moiré patterns nor the positions of the labeled atoms relative to one moiré patterns has changed (by comparing Fig. 6a with Fig. 6c), in other words, it does not produce new configurations. To conclude, the number of micro-status of tBLG is directly proportional to the area of one moiré unit-cell.

Further, we state that the above conclusions also hold true for incommensurate supercells. The difference between incommensurate and commensurate superlattices is that though their geometric moiré patterns are strictly periodic, the positions of atoms located at different moiré unit-cells are not one-to-one correspondence for the incommensurate superlattices. However, we noted that the energy of any atom within each of moiré unit-cells must be within the range of $[\varepsilon_{\mathrm{AB}}, \varepsilon_{\mathrm{AA}}]$ (Fig. 5b), where $\varepsilon_{\mathrm{AB}}$ and $\varepsilon_{\mathrm{AA}}$ represent the atom being above the center and above the vertex of the six-membered ring of the bottom graphene, respectively. Hence, we can choose an energy range $(\varepsilon_i, \varepsilon_i + \mathrm{d}\varepsilon_i]$, where $\varepsilon_i \in [\varepsilon_{\mathrm{AB}}, \varepsilon_{\mathrm{AA}}]$, within which there is exactly one atom in any of the moiré unit-cells (Fig. 5b). Therefore, a one-to-one correspondence can be established between atoms locating in different moiré unit cells, and the number of micro-status of the any twisted bilayer graphene (with incommensurate superlattices) is also proportional to the area of one moiré unit-cell.

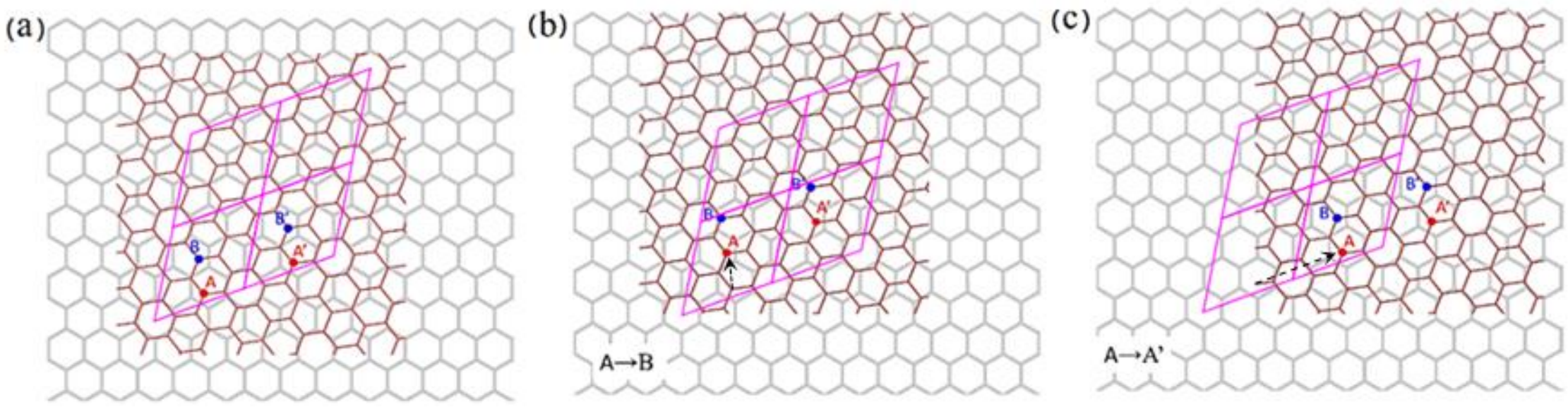

**Fig. 6.** (a)-(c) A graphene flake (brown hexagons) twists on a bottom graphene (grey hexagons), the twist angle is 21.78°. There are 14 atoms in each moiré unit cell. The translational motion of the upper graphene can be represented by the translation of any one of its atoms such as A, B when neglecting the in-plane deformation (rigid approximation).

Considering that the finite size of the top graphene flake makes its boundaries distinguishable and the moiré supercell possesses six axes of symmetry, the total number of the micro-status is 12th power of the number of micro-status determined by the area of one moiré unit-cell. According to the Boltzmann's entropy equation, $S_\theta = k_B \ln \Omega_\theta$, where $\Omega_\theta$ is the number of micro-status of tBLG with a twist angle of $\theta$. Then we have $\Omega_\theta \propto A_\theta^{12}$, where $A_\theta = \sqrt{3}a_m^2/2$ is the area of one moiré unit-cell. We can herein define the entropy change associated with the twist angle as:

$$\Delta S = k_B \ln \frac{\Omega_\theta}{\Omega_0} = 12 k_B \ln \frac{A_\theta}{A_0} 24 k_B \ln \frac{a_\theta}{a} \tag{16}$$

where $A_0 = \sqrt{3}a^2/2$ is the area of one unit-cell at AB-stacked state ($\theta = 0°$). Eq. (16) can be further simplified as $\Delta S = 24 k_B \ln\left(\frac{1}{2}\csc\frac{\theta}{2}\right)$, which is exactly the third term in Eq. (15). We therefore named it as the configuration entropy.

*2.5 The torque and the shear stress during quasi-static rotation of tBLG*

Figure 4 reveals that the configuration entropy originating from the twisting induced disorder plays a dominate role in the Helmholtz free energy (Eq. (16)), therefore, it could also play an important role in twist angle related physical processes, such as superlubric rotation motion. To investigate the contribution of configuration entropy on the interlayer friction of superlubric rotation motion, we considered a square graphene flake with a side length of $L$ and rotated it quasi-statically relative to the center of a hexagon of the fixed bottom layer graphene under torque $M$.

According to the first law of thermodynamics, the energy conservation requires:

$$\mathrm{d}E = \mathrm{d}W + \mathrm{d}Q \tag{17}$$

where $\mathrm{d}E$ is the potential energy change of the system, $\mathrm{d}W = M\mathrm{d}\theta$ is the work done to the system by the torque, and $\mathrm{d}Q = T\mathrm{d}S$ is the heat. Using the relation among the Helmholtz free energy, the internal energy (potential energy here) and the entropy:

$$F = E - TS \tag{18}$$

And substitute Eqs. (15) and (18) into (17), the torque $M$ is obtained as:

$$M = \frac{\partial(\Delta F)}{\partial\theta} = \rho_S L^2 \frac{\mathrm{d}(\Delta u)}{\mathrm{d}\theta} + 12k_B T \cot\frac{\theta}{2} \tag{19}$$

where $\Delta u$ is the difference of interlayer potential energy between the twisted and the AB-stacked bilayer graphene.

On the other hand, by considering the force balance of the upper graphene and assuming a uniform distribution of the interlayer shear stress ($\tau$), we have:

$$M = 8\int_0^{\frac{\pi}{4}}\int_0^{\frac{L}{2\cos\theta}} \tau\, r^2\, \mathrm{d}r\mathrm{d}\theta = \frac{\alpha}{6}L^3\tau \tag{20}$$

where $\alpha = \sqrt{2} + \mathrm{artanh}\frac{\pi}{8}$. According to Eqs. (S20) and (S21), we finally obtain:

$$\tau \triangleq \tau_U + \tau_S = \frac{6\rho_S}{\alpha L}\frac{\mathrm{d}(\Delta u)}{\mathrm{d}\theta} + \frac{72k_B T}{\alpha L^3}\cot\frac{\theta}{2} \tag{21}$$

where $\tau_U = \frac{6\rho_S}{\alpha L}\frac{\mathrm{d}(\Delta u)}{\mathrm{d}\theta}$ and $\tau_S = \frac{72k_B T}{\alpha L^3}\cot\frac{\theta}{2}$, which correspond to the interlayer shear stress originating from the interlayer potential energy and the configuration entropy, respectively.

It is noted that although $\tau_S$ is obtained based on a quasi-static process, the change of entropy induced shear stress is still valid for the estimation of the dynamic friction as long as the in-plane deformation of the system is negligible during sliding, which holds true for the superlubric rotation motion (Dienwiebel et al., 2004; Liu et al., 2012b).

Another remarkable feature of $\tau_S$ is that it can be either the driving force or the resistance force depending on the rotation direction. If the rotation direction is from the incommensurate contact to the AB-stacked state (along the opposite direction of the arrows in Fig. 7a), the friction is a driving force, and conversely, it is the resistance force. This finding can well explain the experimental observation by Feng *et al.* (Feng et al., 2013), where they found that the average sliding distance of graphene nanoflakes is larger at a lower temperature (5 K vs. 77 K). Because $\tau_S$ is proportional to temperature, the higher temperature will result in the higher driving force to rotate, therefore, the shorter time is needed to rotate from incommensurate to AB-stacked states, which leads to a shorter sliding distance.

In addition, we noted that the configuration entropy induced friction is size dependent, $\tau_s \propto 1/L^3$ (Eq. (21)), which tells that the friction at the superlubric rotation will significantly decrease as the size of the top graphene increases. Specifically, the interlayer shear stress can be reduced from $10^0$-$10^1$ MPa at $L$ = 10 nm to $10^0$-$10^2$ Pa at $L$ = 1000 nm at $L$ = 1000 nm (depending on the twist angle) (Fig. 7b).

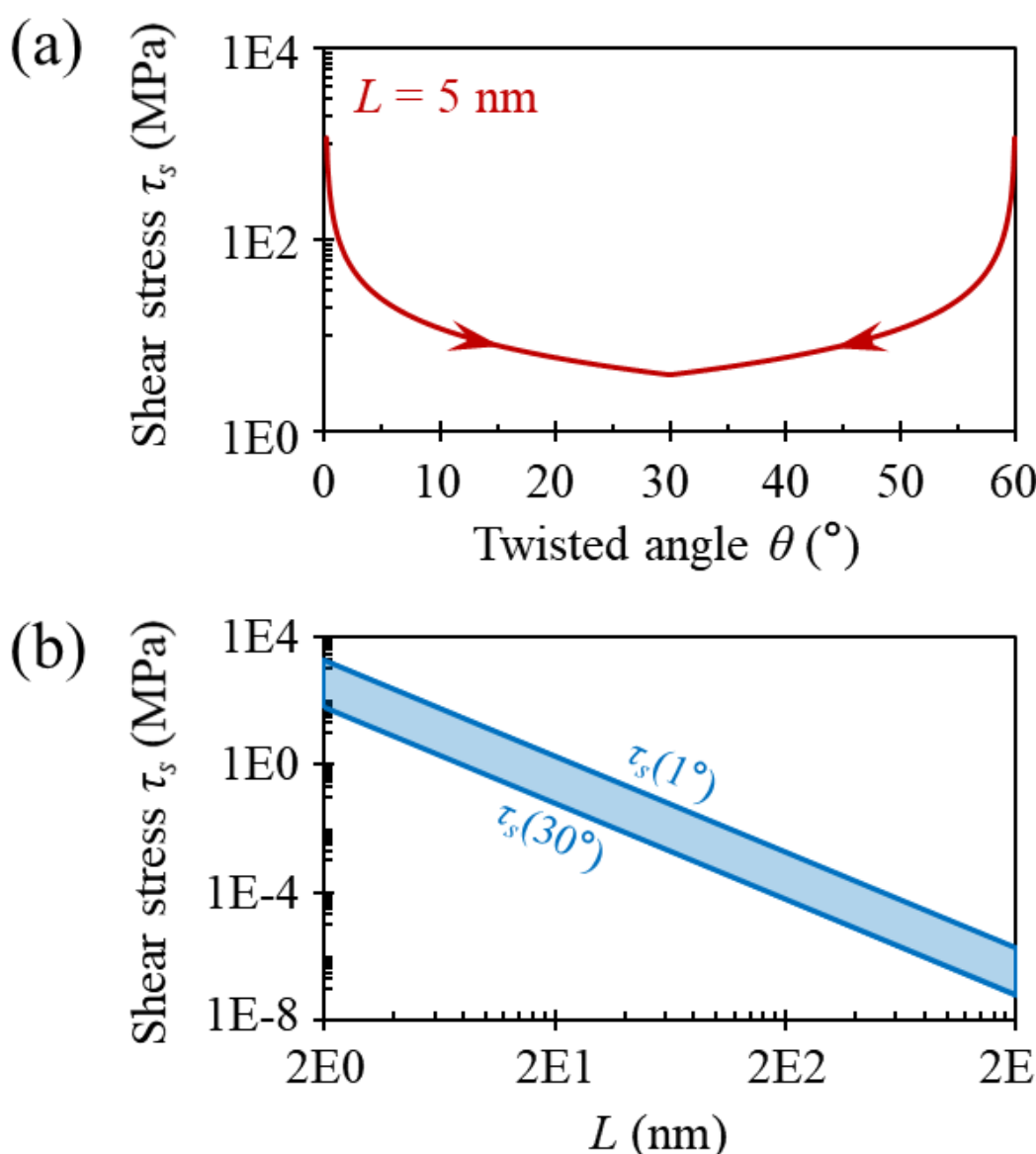

**Fig. 7.** The dependence of the configuration entropy induced interlayer shear stress ($\tau_S$) on the twist angle and the sample size. (a) Typical result for the dependence of $\tau_S$ on the twist angle, where the size of the top graphene flake is set as $L$ = 5 nm. (b) Size dependence of $\tau_S$ during superlubric rotating.

### 3 A dissipation mechanism based on the moiré-dependent deformation

In the above discussions, the friction contributed from the configuration entropy only applies to the cases that rotation motion is involved. As for translational motion, the change of configuration entropy during superlubric sliding is negligible, thus the measured friction at translational motion should originate from other mechanisms (Dienwiebel et al., 2004). Here, we propose a new possible dissipation mechanism based on the finding that the out-of-plane deformation follows the period of moiré supercells (Fig. 1b- d). If the top graphene flake moves along the $x$-direction (e.g. zigzag direction in Fig. 1a) with a constant sliding velocity of $v$, then the accompanying moiré pattern moves with a velocity of $v_m = va_m/a$ along the direction with an angle of $(\pi + \theta)/2$ (Hermann, 2012) with respect to the $x$-axis. Assuming that the moiré level out-of-plane deformation dominates the energy dissipation, we have (Mandelli et al., 2019):

$$\langle f \rangle \cdot v = m\eta \left( 2a_m \langle \Delta z \rangle \frac{v_m}{a\langle \Delta x \rangle} \right)^2 \tag{22}$$

where $\langle f \rangle$ is the average friction at the steady-state sliding process, $\eta$ is the dissipation coefficient, $m$ is the mass of the top graphene flake, $\langle \Delta z \rangle$ and $\langle \Delta x \rangle$ are the amplitude and full-width-at-half-maximum (FWHM) of the out-of-plane deformation, respectively. The out-of-plane deformation of tBLG with different twisted angles was simulated as stated before (see Section 2.3 for details), and the heights of the atoms on the line connecting the centers of two adjacent moiré unit-cells were plotted (Fig. 8). The position along $x$ axis is normalized with the moiré periodicity $a_m$.

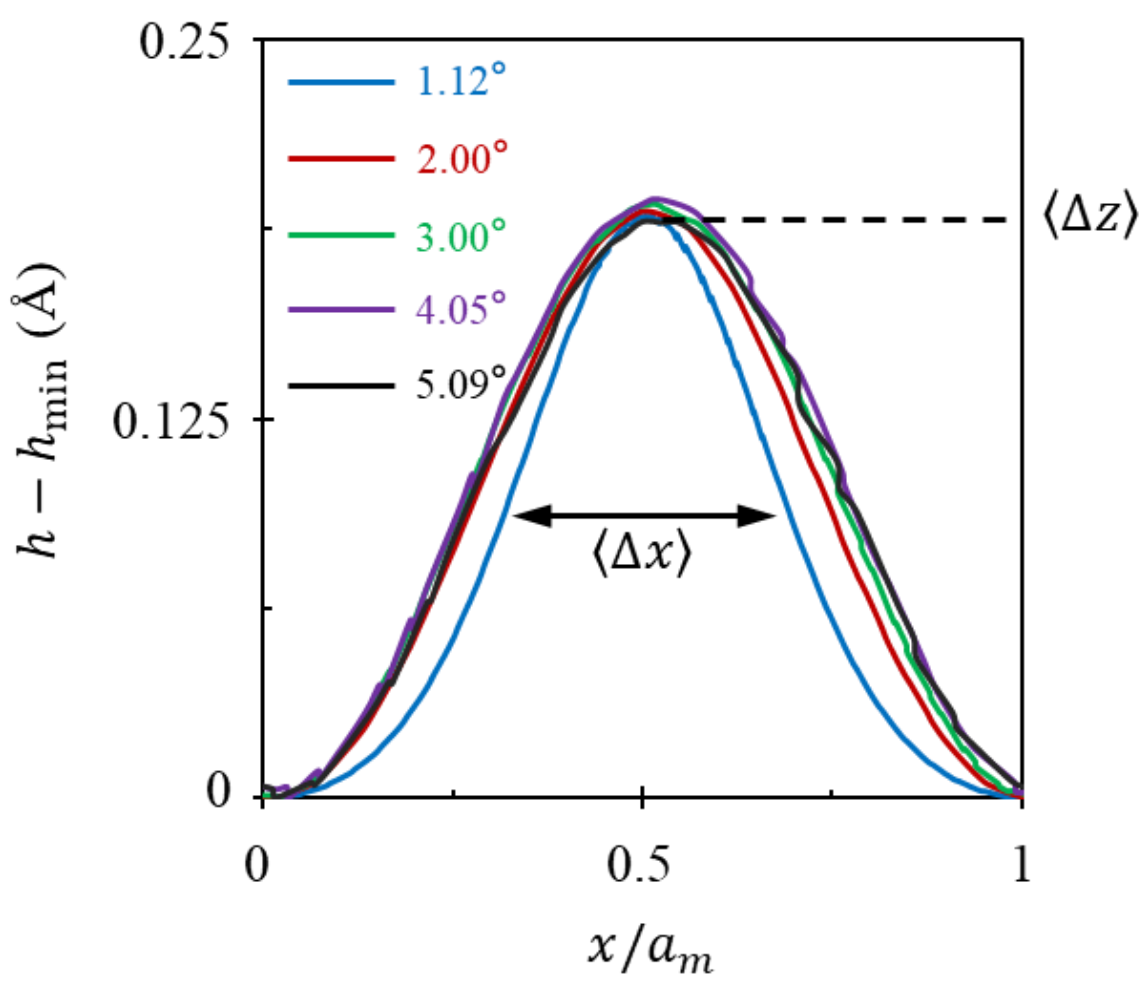

**Fig. 8.** Characters of the out-of-plane deformation of tBLG based on MD simulations.

Using the experimental data in Ref. (Dienwiebel et al., 2004), i.e., $v \sim 30$ nm/s and the friction force of $\langle f \rangle \sim 20$ pN, and the values of $\langle \Delta z \rangle$ and $\langle \Delta x \rangle$ calculated from the MD simulations (Fig. 8), the dissipation coefficient of the superlubric sliding is estimated as $\eta \sim 1$ $\mathrm{ps}^{-1}$, which is the typical value used in most MD simulations (Gigli et al., 2017; Kawai et al., 2016; Leven et al., 2013).

## 4 Closing remarks

To summarize, based on the fact that the out-of-plane deformation in incommensurate bilayer graphene follows the period of moiré supercells, and starting from the classical statistical mechanics, we developed a concise thermodynamic model for twisted bilayer graphene, based on which the analytical expression of configuration entropy is obtained and physically interpreted with the number of micro-status within a moiré supercell. It is found that the configuration entropy plays dominant role to the Helmholtz free energy and the superlubric rotation motion. Besides, the predicted results of this model agree with the experimental observations based on structural superlubricity, such as the observed spontaneous rotation from incommensurate contact to AB-stacked state at finite temperature. Finally, based on the moiré-dependent deformation mode, a dissipation mechanism involved in translational motion of tBLG is proposed. Our work provides the theoretical foundation for incommensurate contact interface and could facilitate twisted bilayer material related applications such as superlubricity.

## Acknowledgments

The authors thank Dr. E. Gao from Wuhan University for scientific discussions. Z. L. acknowledges supports from the starting-up fund of Wuhan University and the support from the Wuhan Science and Technology Bureau of China (No. 2019010701011390). L. S. acknowledges supports from the National Natural Science Foundation of China (No. 11902226). W.O. acknowledges supports from the National Natural Science Foundation of China (No. 12102307) and starting-up fund of Wuhan University.